\documentclass[12pt,a4paper]{article}
\usepackage{amsmath}
\usepackage{graphicx}
\usepackage{amsfonts}
\usepackage{amssymb}

\begin{document}

\author{Remo Garattini\\Universit\`{a} degli Studi di Bergamo, Facolt\`{a} di Ingegneria,\\Viale Marconi, 5, 24044 Dalmine (Bergamo) ITALY\\E-mail: Garattini@mi.infn.it}
\title{A Spacetime Foam Approach to the Schwarzschild-de Sitter Entropy}
\maketitle
\begin{abstract}
The entropy for a black hole in a de Sitter space is approached within the
framework of spacetime foam. A simple model, made by $N$ wormholes in a
semiclassical approximation, is taken under examination to compute the entropy
for such a case. An extension to the extreme case when the black hole and
cosmological horizons are equal is discussed.
\end{abstract}

\section{Introduction}

Black holes have many properties analogous to those of thermodynamics. In
particular, four laws of black holes\cite{BCH} combined with the generalized
second law make up \ a main framework of the black hole thermodynamics. In
these laws, \textit{black hole entropy} is defined as
\begin{equation}
S_{BH}=\frac{A}{4G},
\end{equation}
where $A$ is the area of black hole horizon. This formula is known as
\textit{Bekenstein-Hawking formula}, since the concept of black hole entropy
was first introduced by Bekenstein\cite{J.Bekenstein} as a quantity
proportional to the horizon area and the proportionality coefficient was fixed
by Hawking's discovery of thermal radiation with temperature given by
\begin{equation}
k_{B}T_{BH}=\frac{\hbar\kappa}{2\pi c}. \label{i1}%
\end{equation}
$\kappa$ is the surface gravity of a background black hole. This thermal
radiation and its temperature are called \textit{Hawking radiation} and
\textit{Hawking temperature}, respectively. Let us recall basic properties of
the black hole thermodynamics by taking the example of a one-parameter family
of Schwarzschild black holes, parameterized by the mass $M$. The first law of
thermodynamics\cite{BCH}, in this case is
\begin{equation}
\delta E_{BH}=T_{BH}\delta S_{BH},
\end{equation}
where $E_{BH}$, $S_{BH}$ and $T_{BH}$ are quantities that are identified with
the energy, the entropy and the temperature of a black hole, respectively. The
energy of the black hole is simply given by $E_{BH}=M_{BH}$. A simple relation
exists also for the temperature $T_{BH}$. Hawking showed that a black hole
with surface gravity $\kappa$ emits thermal radiation of a quantum matter
field (which plays the r\^{o}le of a thermometer) at temperature given by Eq.
$\left(  \ref{i1}\right)  $ \cite{S.Hawking}. Since $\kappa=c^{4}/4GM_{BH}$,
it is natural to define the temperature of a Schwarzschild black hole with
mass $M_{BH}$ by
\begin{equation}
k_{B}T_{BH}=\frac{\hbar c^{3}}{8\pi GM_{BH}}. \label{i2}%
\end{equation}
Then from Eqs.$\left(  \ref{i1}\right)  $-$\left(  \ref{i2}\right)  $, we get
\begin{equation}
S_{BH}=\frac{k_{B}c^{3}}{4\hbar G}A+C,
\end{equation}
where $A\equiv16\pi G^{2}M_{BH}^{2}/c^{4}$ is the area of the event horizon
and $C$ is some constant. Since a value of $C$ is not essential in our
discussions, we shall set hereafter $C=0$. This is a special case of the
Bekenstein-Hawking formula. It can be shown that classically the area of the
event horizon cannot decrease in time (the area law\cite{S.W.Hawking2} or the
\textit{second law} of black hole) just as the ordinary thermodynamical
entropy. This observation was the real motivation of introducing a black-hole
entropy\cite{J.Bekenstein}. Moreover, when quantum effects are taken into
account, it is believed that a sum of the black hole entropy and matter
entropy does not decrease (\textit{the generalized second law}). The
\textit{zeroth law} of black hole thermodynamics states that surface gravity
of a Killing horizon is constant throughout the horizon. This is expected for
the Schwarzschild black hole, because it is a static black hole. What is
unexpected is that the same result is valid also for a Kerr black hole, which
is dependent by its temperature, while the surface gravity is not. Of course
we could just check this result but the point is that it is. This supports the
choice of the black-hole temperature. The \textit{third law} does hold in the
sense of Nernst: it is impossible by any process, no matter how idealized, to
reduce \ the surface gravity to zero in a finite sequence by operations
\cite{BCH}. Thermodynamics has a well-established microscopic description: the
quantum statistical mechanics. In the thermodynamical description, information
on each microscopic degree of freedom is lost, and only macroscopic variables
are concerned. However, the number of all microscopic degrees of freedom is
implemented in a macroscopic variable: entropy $S$ is related to the number of
all consistent microscopic states $N$ as
\begin{equation}
S=k_{B}\ln N.
\end{equation}
In analogy, it is expected that there might be a microscopic description of
the black hole thermodynamics, too. In particular, it is widely believed that
the black hole entropy might be related to a number of microscopic states.
This microscopic description seems to require a yet to be developed quantum
theory of gravity. Actually a microscopic derivation of the black hole entropy
was given in superstring theory\cite{GSE,Polchinski,PCJ} by using the
so-called D-brane technology. In this approach, the black hole entropy is
identified with the logarithm of the number of states of massless strings
attached to D-branes, with D-brane configuration and total momentum of the
strings along a compactified direction fixed to be consistent with the
corresponding black hole\cite{SV,Maldacena}. The analysis along this line was
extended to the so-called M-theory\cite{Schwarz}. Recently a different
approach based on a foamy structure of space-time has been proposed
\cite{Remo1,Remo2}. In this approach space-time foam is described by a
collection of $N$ coherent wormholes, whose energy density (Casimir energy) at
its minimum, is
\[
\Delta E_{s}\left(  M\right)  \sim-N_{w}^{2}\frac{V}{64\pi^{2}}\frac
{\Lambda^{4}}{e}.
\]
$\Lambda$ is an U.V. cut-off, $V$ is the volume of the space and $N_{w}$ is
the wormholes number. When we apply the wormhole model to the area, we obtain
the mass quantization of the Schwarzschild black hole, namely\footnote{Units
in which $\hbar=c=k=1$ are used throughout the paper.}
\begin{equation}
S=4\pi M^{2}G=4\pi M^{2}l_{p}^{2}=N\pi\Longrightarrow M=\frac{\sqrt{N}}%
{2l_{p}}.
\end{equation}
A second consequence is that in de Sitter space, the cosmological constant is
quantized in terms of $l_{p}$, i.e.
\begin{equation}
S=\frac{3\pi}{l_{p}^{2}\Lambda_{c}}=\frac{A}{4l_{p}^{2}}=\frac{N4\pi l_{p}%
^{2}}{4l_{p}^{2}}=N\pi\Longrightarrow\frac{3}{l_{p}^{2}N}=\Lambda_{c}.
\end{equation}
In this paper we would like to apply the same wormhole model of spacetime
foam, to compute the entropy of a black hole embedded in a de Sitter space
whose line element is described by the Schwarzschild-de Sitter metric (SdS).
We will also look at its extreme version, the so-called Nariai
metric\cite{Nariai}.\bigskip\ The plan of the paper is the following: in
section \ref{p1}, we will briefly report our model of space-time foam, in
section \ref{p2}, we give a simple example of application of the resulting
discretized (foamy) spacetime to the computation of the entropy in the
Schwarzschild and in de Sitter case; in section \ref{p3}, we discuss the
entropy quantization in the case of the SdS case. We summarize and conclude in
section \ref{p4}.

\section{Constructing the Foam}

\label{p1}

When we try to merge General Relativity with Quantum Field Theory at the
Planck scale, spacetime could be subjected to topology and metric fluctuations
\cite{Wheeler}\footnote{It is interesting to note that there are also
indications on how a foamy spacetime can be tested experimentally\cite{GAC}.}.
Such a fluctuating spacetime is known under the name of ``\textit{spacetime
foam}'' which can be taken as a model for the quantum gravitational vacuum. At
this scale of lengths (or energies) quantum processes like black hole pair
creation could become relevant. To establish if a foamy spacetime could be
considered as a candidate for a Quantum Gravitational vacuum, we can examine
the structure of the effective potential for such a spacetime. There are some
examples showing that flat space cannot be considered as the true ground state
for General Relativity \cite{GPY,HH,CS,Remo}. In the case of Ref.\cite{Remo},
the whole spacetime has been considered as a black hole-anti-black hole pair
formed up by a black hole with positive mass $M$ in the coordinate system of
the observer and an \textit{anti black-hole} with negative mass $-M$ in the
system where the observer is not present. In this way we have an energy
preserving mechanism, because flat space has \textit{zero energy} and the pair
has zero energy too. However, in this case we have not a cosmological
\textit{force} producing the pair: we have only pure gravitational
fluctuations. The black hole-anti-black hole pair has also a relevant
pictorial interpretation: the black hole with positive mass $M$ and the
\textit{anti black-hole} with negative mass $-M$ can be considered the
components of a virtual dipole with zero total energy created by a large
quantum gravitational fluctuation\cite{Modanese}. Note that this is the only
physical process compatible with the energy conservation. The importance of
having the same energy behaviour (\textit{asymptotic}) is related to the
possibility of having a spontaneous transition from one spacetime to another
one with the same boundary condition \cite{Witten}. This transition is a decay
from the false vacuum to the true one\cite{Coleman,Mazur}. However, if we take
account of a pair of neutral black holes living in different universes, there
is no decay and more important no temperature is involved to change from flat
to curved space. To see if this process is realizable we need to compute
quantum corrections to the energy stored in the boundaries. These quantum
corrections are pure gravitational vacuum excitations which can be measured by
the Casimir energy, formally defined as
\begin{equation}
E_{Casimir}\left[  \partial\mathcal{M}\right]  =E_{0}\left[  \partial
\mathcal{M}\right]  -E_{0}\left[  0\right]  , \label{i0}%
\end{equation}
where $E_{0}$ is the zero-point energy and $\partial\mathcal{M}$ is a
boundary. \bigskip We begin to consider the following line element
(Einstein-Rosen bridge) related to a single wormhole
\begin{equation}
ds^{2}=-N^{2}\left(  r\right)  dt^{2}+\frac{dr^{2}}{1-\frac{2MG}{r}}%
+r^{2}\left(  d\theta^{2}+\sin^{2}\theta d\phi^{2}\right)
\end{equation}
We\bigskip\ wish to compute the Casimir-like energy
\[
\Delta E\left(  M\right)  =E\left(  M\right)  -E\left(  0\right)
\]%
\begin{equation}
=\frac{\left\langle \Psi\left|  H^{Schw.}-H^{Flat}\right|  \Psi\right\rangle
}{\left\langle \Psi|\Psi\right\rangle }+\frac{\left\langle \Psi\left|
H_{ql}\right|  \Psi\right\rangle }{\left\langle \Psi|\Psi\right\rangle }.
\end{equation}
by perturbing the three-dimensional spatial metric $g_{ij}=\tilde{g}%
_{ij}+h_{ij}$. $\Delta E\left(  M\right)  $ is computed in a WKB
approximation, by looking at the graviton sector (spin 2 or TT tensor) in a
Schr\"{o}dinger representation with trial wave functionals of the Gaussian
form by means of a variational approach. The Spin-two operator is defined as
\begin{equation}
\left(  \triangle_{2}\right)  _{j}^{a}:=-\triangle\delta_{j}^{a}+2R_{j}^{a}%
\end{equation}
where $\triangle$ is the Laplacian on a Schwarzschild background and
$R_{j}^{a}$ is the mixed Ricci tensor whose components are:
\begin{equation}
R_{j}^{a}=diag\left\{  \frac{-2MG}{r^{3}},\frac{MG}{r^{3}},\frac{MG}{r^{3}%
}\right\}  .
\end{equation}
The total energy at one loop, i.e., the classical term plus the stable and
unstable modes respectively, is
\[
\Delta E_{q.l.}+\Delta E_{s}+\Delta E_{u}%
\]
where $\Delta E_{q.l.}$ is the quasilocal energy. For symmetric boundary
conditions with respect to the bifurcation surface $S_{0}$ (such as this case
$E_{q.l.}=E_{+}-E_{-}=0$. When the boundaries go to spatial infinity $E_{\pm
}=M_{ADM}$. The \noindent Stable modes contribution is
\begin{equation}
\Delta E_{s}=-\frac{V}{32\pi^{2}}\left(  \frac{3MG}{r_{0}^{3}}\right)  ^{2}%
\ln\left(  \frac{r_{0}^{3}\Lambda^{2}}{3MG}\right)  .
\end{equation}
$\Lambda$ is a cut-off to keep under control the $UV$ divergence, we can think
that $\Lambda\leq m_{p}$. For the unstable sector, there is only \textbf{one
eigenvalue in S-wave.} This is in agreement with Coleman arguments on quantum
tunneling: the presence of a unique negative eigenvalue in the second order
perturbation is a signal of a passage from a false vacuum to a true vacuum.
The Rayleigh-Ritz method joined to a numerical integration technique gives
$E^{2}=-.\,\allowbreak175\,41/\left(  MG\right)  ^{2}$\cite{Remo}, to be
compared with the value $E^{2}=-.\,\allowbreak19/\left(  MG\right)  ^{2}$ of
Ref.\cite{GPY}.\bigskip\ How to eliminate the instability? We consider $N_{w}%
$\ coherent wormholes (i.e., non-interacting) in a semiclassical approximation
and assume that there exists a covering of $\Sigma$\ such that $\Sigma
=\cup_{i=1}^{N_{w}}\Sigma_{i}$, with $\Sigma_{i}\cap\Sigma_{j}=\emptyset
$\ when $i\neq j$. Each $\Sigma_{i}$\ has the topology $S^{2}\times R^{1}%
$\ with boundaries $\partial\Sigma_{i}^{\pm}$\ with respect to each
bifurcation surface. On each surface $\Sigma_{i}$, quasilocal energy is zero
because we assume that on each copy of the single wormhole there is symmetry
with respect to each bifurcation surface. Thus the total energy for the
collection is $E_{|2}^{tot}=N_{w}H_{|2}$ and the total trial wave functional
is the product of $N_{w}$ t.w.f.
\begin{equation}
\Psi_{tot}^{\perp}=\Psi_{1}^{\perp}\otimes\Psi_{2}^{\perp}\otimes\ldots
\ldots\Psi_{N_{w}}^{\perp}%
\end{equation}

By repeating the same calculations done for the single wormhole for the
N$_{w}$\ wormhole system, we obtain

\begin{description}
\item [ a)]The total Casimir energy (stable modes), at its minimum, is
\[
\Delta E_{s}\left(  M\right)  \sim-N_{w}^{2}\frac{V}{64\pi^{2}}\frac
{\Lambda^{4}}{e}.
\]
The minimum does not correspond to flat space $\rightarrow$ $\Delta
E_{s}\left(  M\right)  \neq0.$

\item[ b)] The initial boundary located at $R_{\pm}$ will be reduced to
$R_{\pm}/N_{w}.$

\item[ c)] Since the boundary is reduced there exists a critical radius
$\rho_{c}=1.\,\allowbreak113\,4$ such that : $\forall N\geq N_{w_{c}}%
\ \exists$ $r_{c}$ $s.t.$ $\forall\ r_{0}\leq r\leq r_{c},\ \sigma\left(
\Delta_{2}\right)  =\emptyset$. This means that the system begins to be
stable\cite{Remo1,Remo2}. To be compared with the value $\rho_{c}=1.445$
obtained by B. Allen in Ref.\cite{Allen}.
\end{description}

\section{Area Spectrum, Entropy and the Cosmological constant}

\label{p2}

Bekenstein has proposed that a black hole does have an entropy proportional to
the area of its horizon $S_{bh}=const\times A_{hor}$\cite{J.Bekenstein}. In
natural units one finds that the proportionality constant is set to
$1/4G=1/4l_{p}^{2}$, so that the entropy becomes $S=A/4G=A/4l_{p}^{2}.$
Another proposal always made by Bekenstein is the quantization of the area for
nonextremal black holes $a_{n}=\alpha l_{p}^{2}\left(  n+\eta\right)  $%
\qquad$\eta>-1$\qquad$n=1,2,\ldots$ The area is measured by the quantity
\begin{equation}
A\left(  S_{0}\right)  =\int_{S_{0}}d^{2}x\sqrt{\sigma}.
\end{equation}
We would like to evaluate the mean value of the area
\begin{equation}
A\left(  S_{0}\right)  =\frac{\left\langle \Psi_{F}\left|  \hat{A}\right|
\Psi_{F}\right\rangle }{\left\langle \Psi_{F}|\Psi_{F}\right\rangle }%
=\frac{\left\langle \Psi_{F}\left|  \widehat{\int_{S_{0}}d^{2}x\sqrt{\sigma}%
}\right|  \Psi_{F}\right\rangle }{\left\langle \Psi_{F}|\Psi_{F}\right\rangle
},
\end{equation}
computed on the foam state
\begin{equation}
\left|  \Psi_{F}\right\rangle =\Psi_{1}^{\perp}\otimes\Psi_{2}^{\perp}%
\otimes\ldots\ldots\Psi_{N_{w}}^{\perp}.
\end{equation}
Consider $\sigma_{ab}=\bar{\sigma}_{ab}+\delta\sigma_{ab}\bar{\sigma}_{ab} $
is such that $\int_{S_{0}}d^{2}x\sqrt{\bar{\sigma}}=4\pi\bar{r}^{2}$ and
$\bar{r}$ is the radius of $S_{0}$%
\begin{equation}
A\left(  S_{0}\right)  =\frac{\left\langle \Psi_{F}\left|  \hat{A}\right|
\Psi_{F}\right\rangle }{\left\langle \Psi_{F}|\Psi_{F}\right\rangle }=4\pi
\bar{r}^{2}.
\end{equation}
Suppose to consider the mean value of the area $A$ computed on a given
\textit{macroscopic} fixed radius $R$. On the basis of our foam model, we
obtain $A=\bigcup\limits_{i=1}^{N}A_{i}$, with $A_{i}\cap A_{j}=\emptyset$
when $i\neq j$. Thus
\begin{equation}
A=4\pi R^{2}=\sum\limits_{i=1}^{N}A_{i}=\sum\limits_{i=1}^{N}4\pi\bar{r}%
_{i}^{2}.
\end{equation}
When $\bar{r}_{i}\rightarrow l_{p}$, $A_{i}\rightarrow A_{l_{P}}$
and\cite{Remo2}
\begin{equation}
A=NA_{l_{P}}=N4\pi l_{p}^{2}\Longrightarrow S=\frac{A}{4l_{p}^{2}}=\frac{N4\pi
l_{p}^{2}}{4l_{p}^{2}}=N\pi. \label{p21}%
\end{equation}
Thus the macroscopic area is represented by $N$\ microscopic areas of the
Planckian size. In this sense we will claim that the area is quantized. The
first consequence is the mass quantization of the Schwarzschild black hole,
namely
\begin{equation}
S=4\pi M^{2}G=4\pi M^{2}l_{p}^{2}=N\pi\Longrightarrow M=\frac{\sqrt{N}}%
{2l_{p}}.
\end{equation}
To be compared with Refs.\cite{Ahluwalia,Hod,Kastrup,Makela,Mazur1,Vaz}. A
second consequence is that in de Sitter space, the cosmological constant is
quantized in terms of $l_{p}$, i.e.\cite{JGB,Zizzi}
\begin{equation}
S=\frac{3\pi}{l_{p}^{2}\Lambda_{c}}=\frac{A}{4l_{p}^{2}}=\frac{N4\pi l_{p}%
^{2}}{4l_{p}^{2}}=N\pi\Longrightarrow\frac{3}{l_{p}^{2}N}=\Lambda_{c}.
\label{p22}%
\end{equation}
It is possible to give an estimate of the total amount of Planckian wormholes
needed to fill the space beginning from the Planck era $\left(  \Lambda
\sim\left(  10^{16}-10^{18}GeV\right)  ^{2}\right)  $ up to the space in which
we now live $\Lambda\leq\left(  10^{-42}GeV\right)  ^{2}$.
\begin{equation}
\frac{1}{N}10^{38}GeV^{2}=10^{-84}GeV^{2}\rightarrow N=10^{122},
\end{equation}
in agreement with the observational data $\Lambda_{c}\lesssim10^{-122}%
l_{P}^{-2}$ coming from the Friedmann-Robertson-Walker cosmology constraining
the cosmological constant\cite{MVisser}.

\section{Entropy for black holes in de Sitter space}

\label{p3}The Schwarzschild-de Sitter metric (SdS) is defined as
\begin{equation}
ds^{2}=-\left(  1-\frac{2MG}{r}-\frac{\Lambda_{c}}{3}r^{2}\right)
dt^{2}+\left(  1-\frac{2MG}{r}-\frac{\Lambda_{c}}{3}r^{2}\right)  ^{-1}%
dr^{2}+r^{2}d\Omega^{2}. \label{3aa}%
\end{equation}
For $\Lambda_{c}=0$ the metric becomes
\begin{equation}
ds^{2}=-\left(  1-\frac{2MG}{r}\right)  dt^{2}+\left(  1-\frac{2MG}{r}\right)
^{-1}dr^{2}+r^{2}d\Omega^{2}%
\end{equation}
and it describes the Schwarzschild metric, while for $M=0$, we obtain
\begin{equation}
ds^{2}=-\left(  1-\frac{\Lambda_{c}}{3}r^{2}\right)  dt^{2}+\left(
1-\frac{\Lambda_{c}}{3}r^{2}\right)  ^{-1}dr^{2}+r^{2}d\Omega^{2},
\end{equation}
namely the de Sitter metric (dS). The gravitational potential $g_{00}\left(
r\right)  $ of $\left(  \ref{3aa}\right)  $ admits three real roots. One is
negative and it is located at
\begin{equation}
r_{-}=\frac{2}{\sqrt{\Lambda_{c}}}\cos\left(  \frac{\theta}{3}+\frac{2\pi}%
{3}\right)  ,
\end{equation}
while the other ones are associated to the black hole and cosmological
horizons respectively located at
\begin{equation}
r_{+}=\frac{2}{\sqrt{\Lambda_{c}}}\cos\left(  \frac{\theta}{3}\right)
,\ r_{++}=\frac{2}{\sqrt{\Lambda_{c}}}\cos\left(  \frac{\theta}{3}+\frac{4\pi
}{3}\right)  , \label{3ab}%
\end{equation}
where
\begin{equation}
\cos\theta=-3MG\sqrt{\Lambda_{c}}, \label{3ac}%
\end{equation}
with the condition
\begin{equation}
0\leq9\left(  MG\right)  ^{2}\Lambda_{c}\leq1. \label{3a}%
\end{equation}
Eq.$\left(  \ref{3ac}\right)  $ implies that $\theta$ $\in\left[  \frac{\pi
}{2},\frac{3\pi}{2}\right]  $. In this interval $r_{+}$ is a monotonic
decreasing function of $\theta$, while $r_{++}$ is a monotonic increasing one
with
\begin{equation}
\left\{
\begin{array}
[c]{c}%
r_{+}^{\downharpoonright}\in\left[  0,\sqrt{3/\Lambda_{c}}\right] \\
r_{++}^{\upharpoonright}\in\left[  0,\sqrt{3/\Lambda_{c}}\right]  .
\end{array}
\right.
\end{equation}
They have a common value when $r_{+}=r_{++}=1/\sqrt{\Lambda_{c}}$, where
$9\left(  MG\right)  ^{2}\Lambda=1$ and $\theta=\pi$. This means that the
cosmological horizon and the black hole horizon have merged. The relation
between the three roots is
\begin{equation}
r_{-}+r_{+}+r_{++}=0
\end{equation}
and furthermore
\begin{equation}
\left\{
\begin{array}
[c]{c}%
3/\Lambda_{c}=r_{+}^{2}+r_{+}r_{++}+r_{++}^{2}\\
6Ml_{p}^{2}/\Lambda_{c}=\left(  r_{+}r_{++}\right)  \left(  r_{+}%
+r_{++}\right)  .
\end{array}
\right.  \label{3b}%
\end{equation}
The gravitational entropy in the SdS case is
\begin{equation}
S=\frac{A_{bh}+A_{c}}{4l_{p}^{2}}=\frac{\pi}{l_{p}^{2}}\left(  r_{+}%
^{2}+r_{++}^{2}\right)  , \label{3c}%
\end{equation}
namely it is the sum of the black hole and cosmological entropy\cite{KT}. By
means of Eqs.$\left(  \ref{3ab}\right)  $ one gets,
\begin{equation}
S=\frac{4\pi}{\Lambda_{c}l_{p}^{2}}\left(  \cos^{2}\left(  \frac{\theta}%
{3}\right)  +\cos^{2}\left(  \frac{\theta}{3}+\frac{4\pi}{3}\right)  \right)
=\frac{4\pi}{\Lambda_{c}l_{p}^{2}}c\left(  \theta\right)  .
\end{equation}
For $\theta\in\left[  \frac{\pi}{2},\frac{3\pi}{2}\right]  $, $c\left(
\theta\right)  \in\left[  \frac{1}{2},\frac{3}{4}\right]  $ and the entropy is
bounded by
\begin{equation}
\frac{2\pi}{\Lambda_{c}l_{p}^{2}}\leq S\leq\frac{3\pi}{\Lambda_{c}l_{p}^{2}}.
\label{p31}%
\end{equation}
The lower bound of inequality $\left(  \ref{p31}\right)  $ corresponds to the
entropy of the Nariai metric, whose (Lorentzian) line element is
\begin{equation}
ds^{2}=-\left(  1-\Lambda p^{2}\right)  dt^{2}+\left(  1-\Lambda p^{2}\right)
^{-1}dp^{2}+\frac{1}{\Lambda}d\Omega^{2},
\end{equation}
Thus the entropy in the SdS case has an upper bound represented by the de
Sitter entropy and a lower bound represented by the Nariai entropy. By means
of Eq.$\left(  \ref{p22}\right)  $, the Nariai entropy is
\begin{equation}
S=\frac{2\pi}{\Lambda_{c}l_{p}^{2}}=\frac{2\pi N}{3}%
\end{equation}
and the relative black hole mass is
\begin{equation}
M=\frac{\left(  r_{+}r_{++}\right)  \left(  r_{+}+r_{++}\right)  }{2l_{p}%
^{2}\left(  r_{+}^{2}+r_{+}r_{++}+r_{++}^{2}\right)  },
\end{equation}
where we have used Eqs.$\left(  \ref{3b}\right)  $ to express the mass and the
cosmological constant in terms of the roots of the gravitational potential
$g_{00}\left(  r\right)  $. When $r_{+}=r_{++}=\bar{r}$, the black hole mass
is equal to the ``\textit{cosmological mass}'' $M_{c}$
\begin{equation}
M_{c}=M=\frac{2\bar{r}^{2}\bar{r}}{6l_{p}^{2}\bar{r}^{2}}=\frac{\bar{r}%
}{3l_{p}^{2}}=\frac{1}{3l_{p}^{2}}\sqrt{\frac{1}{\Lambda_{c}}},
\end{equation}
i.e. Eq. $\left(  \ref{3a}\right)  $ for $\theta=\pi$. Recalling Eq.$\left(
\ref{p22}\right)  $ we obtain
\begin{equation}
M=\sqrt{\frac{1}{9l_{p}^{4}\Lambda_{c}}}=\sqrt{\frac{N}{27l_{p}^{2}}}%
\qquad\Longrightarrow\qquad M=\frac{\sqrt{N}}{3\sqrt{3}l_{p}}.
\end{equation}
Therefore the black hole mass is bounded by
\begin{equation}
0\leq M\leq\frac{1}{3l_{p}^{2}}\sqrt{\frac{1}{\Lambda_{c}}}=\frac{\sqrt{N}%
}{3\sqrt{3}l_{p}}.
\end{equation}
Note that the black hole mass in the Nariai case is lower than the
Schwarzschild case: this is the effect of having a spacetime with a
\textit{positive} cosmological constant which describes a $S^{3}$ topology.
This means that also the black hole radius cannot exceed the cosmological
radius. Thus the black hole mass has an upper bound deriving from the extreme case.

\section{Conclusions}

\label{p4}On the basis of the model of spacetime foam described in section
\ref{p2}, in this paper an attempt to compute the entropy for the
Schwarzschild-de Sitter metric has been performed including also its extreme
case or Nariai metric. It is known that semiclassically, one can assign a
probability measure that leads to computing the logarithm of the number of
microstates as in the case of a thermal system with an entropy
\begin{equation}
S=\log P.
\end{equation}
The SdS case entropy is described by Eq.$\left(  \ref{3c}\right)  $. The
associated probability is
\begin{equation}
P_{SdS}\simeq\exp S=\exp\frac{\pi}{l_{p}^{2}}\left(  r_{+}^{2}+r_{++}%
^{2}\right)  =\exp\frac{4\pi}{\Lambda_{c}l_{p}^{2}}c\left(  \theta\right)  .
\end{equation}
This value interposes two special cases, as seen. The first is the de Sitter
case where $M=0$ and $S=3\pi/l_{P}^{2}\Lambda_{c}$, i.e.
\begin{equation}
P_{dS}\simeq\exp\left(  \frac{3\pi}{l_{P}^{2}\Lambda_{c}}\right)  =\exp\pi N
\end{equation}
and the second is the Nariai case where $M=M_{c}$ and $S=2\pi/l_{P}^{2}%
\Lambda_{c}$ with probability
\begin{equation}
P_{N}\simeq\exp\left(  \frac{2\pi}{l_{P}^{2}\Lambda_{c}}\right)  =\exp
\frac{2\pi N}{3}.
\end{equation}
Thus the probability is an exponentially decreasing function in terms of the
mass parameter and
\begin{equation}
P_{N}\leq P_{SdS}\leq P_{dS}\qquad\Longleftrightarrow\qquad\exp\frac{2\pi
N}{3}\leq\exp\frac{4\pi N}{3}c\left(  \theta\right)  \leq\exp\pi N,
\label{p43}%
\end{equation}
where we have expressed the probability in terms of the wormholes number. As
expected the presence of the cosmological constant modifies the property of
the black hole mass and consequently of its horizon. This modification is
given by directly comparing the Schwarzschild metric and the Schwarzschild-de
Sitter metric. Moreover from Eq.$\left(  \ref{p43}\right)  $, we realize that
the de Sitter space has the best probability to be realized when compared with
the SdS or Nariai spaces. Nevertheless, we can see that the SdS space has a
probability that can be driven close to the de Sitter probability. Moreover,
when we compare $P_{SdS}$ with $P_{N}$ we can see that the SdS space has a
major probability with respect to the extreme space, namely the Nariai space.
This conclusion seems to be in conflict with the request that only regular
Euclidean Einstein solutions close to the horizon have to be considered; in
this case only the Nariai solution. A possibility that can be investigated to
better understand this situation is given by the computation of the
Casimir-like energy for the SdS space with the de Sitter space as a reference
space. This has been done for the Schwarzschild space with flat space as a
reference space in Ref.\cite{Remo} and it has led to the N-wormhole
approximation of the foam we have used to compute the entropy for the SdS and
Nariai spaces. The same steps can be repeated to better understand the process
of black hole pair creation in presence of a cosmological term and its
consequences on the foam structure.

\vglue0.5cm \textbf{\noindent References and Notes} \vglue0.4cm

\begin{list}{\arabic{enumi}.}{\usecounter{enumi}\setlength{\itemsep}{3pt}
\settowidth{\labelwidth}{2em}\sloppy}
\bibitem{BCH}  J.M. Bardeen, B. Carter and S.W. Hawking, Comm. Math. Phys.
\textbf{31}, 161 (1973).
\bibitem{J.Bekenstein}  J. Bekenstein, Phys. Rev. \textbf{D7} (1973) 2333.
\bibitem{S.Hawking}  S.W. Hawking, Comm. Math. Phys. \textbf{43} (1975) 190.
\bibitem{S.W.Hawking2}  S.W. Hawking, Phys. Rev. Lett. \textbf{26} (1971)
1344.
\bibitem{GSE}  M.B. Green, J.H. Schwartz and E. Witten, \textit{Superstring
theory} (Cambridge University Press, 1987).
\bibitem{Polchinski}  J. Polchinski, \textit{Superstring theory} (Cambridge
University Press, 1998).
\bibitem{PCJ}  J. Polchinski, S. Chaudhuri and C.V. Johnson, ``Notes on
D-Branes'', hep-th/9602052.
\bibitem{SV}  A. Strominger and C.\ Vafa, Phys. Lett. B \textbf{379} (1996),
99.
\bibitem{Maldacena}  J.M. Maldacena, ``Black Holes in string Theory'',
hep-th/9607235.
\bibitem{Schwarz}  J.H. Schwarz, in \textit{String Theory, Gauge Theory and
Quantum Gravity}, Proceedings of the Spring School, Trieste, Italy, 1996,
edited by R. Dijkgraaf et al. [Nucl. Phys. B (Proc. Suppl.) \textbf{55B}, 1
(1997)].
\bibitem{Remo1}  R. Garattini - Phys.Lett. \textbf{B 446} (1999) 135,
hep-th/9811187.
\bibitem{Remo2}  R. Garattini - Phys. Lett. \textbf{B 459} (1999) 461,
hep-th/9906074.
\bibitem{Nariai}  S. Nariai, \textit{On some static solutions to Einstein's
gravitational field equations in a spherically symmetric case. }Science
Reports of the Tohoku University, \textbf{34} (1950) 160; S. Nariai, \textit
{%
On a new cosmological solution of Einstein's field equations of gravitation.
}Science Reports of the Tohoku University, \textbf{35} (1951) 62.
\bibitem{Wheeler}  J.A. Wheeler, Ann. Phys. 2, 604 (1957).
\bibitem{GAC}  G. Amelino-Camelia, Nature \textbf{398} (1999) 216,
gr-qc/9808029; G. Amelino-Camelia, Nature \textbf{393} (1999) 763,
astro-ph/9712103; G. Amelino-Camelia, \textit{Gravity-wave interferometers
as probes of a low-energy effective quantum gravity}, gr-qc/9903080.
\bibitem{GPY}  D.J. Gross, M.J. Perry and L.G. Yaffe, Phys. Rev. D\textbf{\
25}, (1982) 330.
\bibitem{HH}  J.B. Hartle and G.T. Horowitz, Phys. Rev. \textbf{D}%
\textbf{24%
}, (1981) 257.
\bibitem{CS}  L. Crane and L. Smolin, Nucl. Phys. \textbf{B} (1986) 714.
\bibitem{Remo}  R. Garattini, Int. J. Mod. Phys. A 18 (1999) 2905,
gr-qc/9805096.
\bibitem{Modanese}  G. Modanese, Phys.Lett. \textbf{B 460} (1999) 276
\bibitem{Witten}  E. Witten, Nucl. Phys. B \textbf{195 (}1982) 481.
\bibitem{Coleman}  S. Coleman, Nucl. Phys. B \textbf{298} (1988) 178.
\bibitem{Mazur}  P.O. Mazur, Mod. Phys. Lett \textbf{A 4}, (1989) 1497.
\bibitem{Allen}  B. Allen, Phys. Rev. D \textbf{30} (1984) 1153.
\bibitem{Ahluwalia}  D.V. Ahluwalia, Int.J.Mod.Phys. D \textbf{8} (1999)
651, astro-ph/9909192.
\bibitem{Hod}  S. Hod, Phys. Rev. Lett. \textbf{81}, 4293 (1998),
gr-qc/9812002.
\bibitem{Kastrup}  H.A. Kastrup, Phys.Lett. \textbf{B 413} (1997) 267,
gr-qc/9707009; H.A. Kastrup, Phys.Lett. \textbf{B 419} (1998) 40,
gr-qc/9710032.
\bibitem{Makela}  J. M\"{a}kel\"{a}, gr-qc/9602008; J. M\"{a}kel\"{a},
Phys.Lett. \textbf{B 390} (1997) 115.
\bibitem{Mazur1}  P.O. Mazur, Acta Phys.Polon. \textbf{27} (1996) 1849,
hep-th/9603014.
\bibitem{Vaz}  C. Vaz and L. Witten, Phys.Rev. \textbf{D 60} (1999) 024009,
gr-qc/9811062.
\bibitem{JGB}  J. Garcia-Bellido, \textit{QUANTUM BLACK HOLES},
hep-th/9302127.
\bibitem{Zizzi}  P. A. Zizzi, Int.J.Theor.Phys. \textbf{38} (1999) 2331,
hep-th/9808180.
\bibitem{MVisser}  M. Visser, \textit{Lorentzian Wormholes} (AIP Press, New
York, 1995) 64.
\bibitem{KT}  D. Kastor and J. Traschen, Class.Quant.Grav. \textbf{13}
(1996) 2753.
\end{list}
\end{document}